\documentstyle[aps,epsf,preprint,floats]{revtex}

\newcommand{\lsim}{\mathrel{\raisebox{-.6ex}{$\stackrel{\textstyle<}{\sim}$}}}
\newcommand{\gsim}{\mathrel{\raisebox{-.6ex}{$\stackrel{\textstyle>}{\sim}$}}}

\def\dmatm{\delta m^2_{\rm atm}}

\def\Pej{P_{ej}}

\def\mb{m^2_{\beta}}

\begin{document}

\tightenlines
\catcode`@=11
\def\references{%
\ifpreprintsty
\bigskip\bigskip
\hbox to\hsize{\hss\large \refname\hss}%
\else
\vskip24pt
\hrule width\hsize\relax
\vskip 1.6cm
\fi
\list{\@biblabel{\arabic{enumiv}}}%
{\labelwidth\WidestRefLabelThusFar  \labelsep4pt %
\leftmargin\labelwidth %
\advance\leftmargin\labelsep %
\ifdim\baselinestretch pt>1 pt %
\parsep  4pt\relax %
\else %
\parsep  0pt\relax %
\fi
\itemsep\parsep %
\usecounter{enumiv}%
\let\p@enumiv\@empty
\def\theenumiv{\arabic{enumiv}}%
}%
\let\newblock\relax %
\sloppy\clubpenalty4000\widowpenalty4000
\sfcode`\.=1000\relax
\ifpreprintsty\else\small\fi
}
\catcode`@=12

\preprint{\font\fortssbx=cmssbx10 scaled \magstep2
\hbox to \hsize{
\hbox{\fortssbx University of Wisconsin - Madison}\hfill
$\vcenter{\tighten
\hbox{\bf MADPH-98-1074}
\hbox{\bf VAND-TH-98-13}
\hbox{\bf AMES-HET-98-11}
\hbox{August 1998}}$}}

\title{\vspace*{.5in}
Inferred 4.4 eV Upper Limits on the\\ Muon- and Tau-Neutrino Masses}

\author{V. Barger$^1$, T.J. Weiler$^2$, and K. Whisnant$^3$}

\address{
$^1$Department of Physics, University of Wisconsin, Madison, WI 53706, USA\\
$^2$Department of Physics and Astronomy, Vanderbilt University, Nashville, TN
37235, USA\\
$^3$Department of Physics and Astronomy, Iowa State University, Ames, IA 50011,
USA}

\maketitle
\thispagestyle{empty}

\begin{abstract}
By combining experimental constraints from atmospheric and solar neutrino
oscillations and the tritium beta decay endpoint, we infer upper limits of
4.4~eV on the $\nu_\mu$ and $\nu_\tau$ masses, if the universe consists
of three neutrinos. For hierarchical mass spectra $m_3 \gg m_1, m_2$
or $m_3 \simeq m_2 \gg m_1$ we infer that $m_{\nu_\alpha} \lsim 0.08$~eV for
$\alpha = e, \mu, \tau$. In a four neutrino universe, and assuming neutrino
oscillations also account for the LSND experimental results,
$m_{\nu_\alpha}~\lsim~5.4$~eV. We also obtain lower limits on the masses.

\end{abstract}

\newpage

The quest to learn whether neutrinos are massive has been long and arduous.
The present laboratory endpoint measurements only give upper
limits\cite{partprop}
\begin{equation}
m_{\nu_e} < 4.4{\rm\ eV}\,, \quad m_{\nu_\mu} < 170 {\rm\ keV}\,, \quad
m_{\mu_\tau} < 18.2 {\rm\ MeV}
\end{equation}
that are not very restrictive in the $\nu_\mu$ and $\nu_\tau$ cases.
However, atmospheric\cite{atmos} and solar\cite{solar,bahcall} oscillation
experiments have recently
obtained positive results, but oscillation experiments
measure only mass-squared differences $\delta m^2_{ij} \equiv m_i^2 -
m_j^2$, leaving the overall mass scale unknown.
Phenomenologically, we have the two inequalities
\footnote{The explanation of the atmospheric data requires
$\nu_\mu\to\nu_\tau$ oscillations (or $\nu_\mu\to\nu_s$ oscillations to a sterile neutrino $\nu_s$) with mass--squared difference\cite{atmos}
$0.5\times 10^{-3} \lsim \delta m^2_{\rm atm} \lsim 6\times 10^{-3}\rm\ eV^2$
and mixing angle $\sin^2 2\theta_{\rm atm} \gsim 0.8$.
The solar neutrino anomaly requires a mass--squared difference of either
$\delta m^2_{sun} \approx 3.5\times 10^{-6}\mbox{--}1\times 10^{-5}$~eV$^2$
with small-angle vacuum mixing
$\sin^2 2\theta_{\rm sun} = 1.5\times 10^{-3}\mbox{--}1.0\times 10^{-2}$
or a mass--squared difference of
$\delta m^2_{sun} \approx 1\times 10^{-10}$~eV$^2$ with large-angle vacuum
mixing $\sin^2 2\theta_{\rm sun} \gsim 0.6$\cite{solar,bahcall}.}
\begin{equation}
\delta m^2_{\rm sun} \ll \delta m^2_{\rm atm} \ll m^2_\beta \,,
\end{equation}
where $m_\beta = 4.4$~eV is the upper limit on the effective $\nu_e$ mass from
tritium beta decay\cite{belesev}.
We show in this Letter that these constraints may be combined to place an upper
bound of $m_\beta$ on the $\nu_\mu$ and $\nu_\tau$ masses in a three-neutrino
universe. To also accomodate the indications for $\nu_\mu \rightarrow
\nu_e$ oscillations in the LSND experiment\cite{LSND} with one sterile
and three active neutinos\cite{caldwell}, we obtain an upper bound
of 5.4~eV on all four neutrino masses.

\begin{figure}
\centering\leavevmode
\epsfxsize=5.5in\epsffile{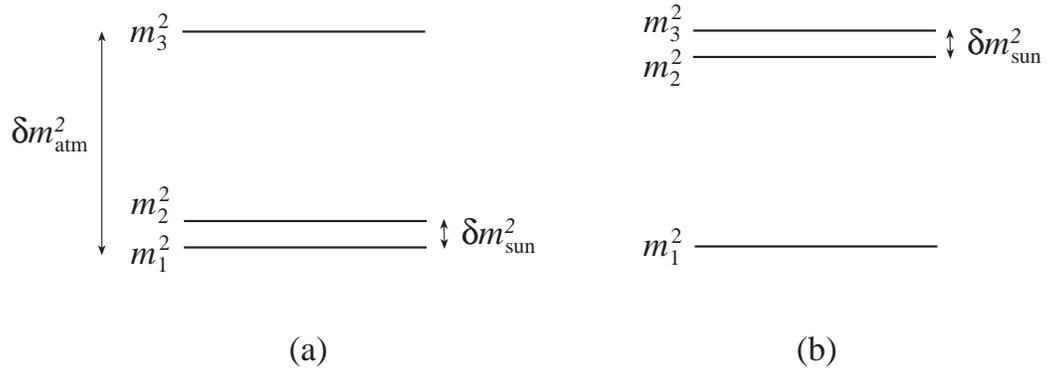}

\bigskip
\caption[]{\label{fig1} The two possible mass scenarios if the atmospheric
and solar neutrino data are described by the oscillations of three
neutrinos.}
\end{figure}

The flavor eigenstates $\nu_e, \nu_\mu, \nu_\tau$ in a three-neutrino
universe are related to the mass
eigenstates $\nu_1, \nu_2, \nu_3$ by a unitarity transformation
\begin{equation}
\left(\begin{array}{c} \nu_e \\ \nu_\mu \\ \nu_\tau \end{array}\right) =
\left(\begin{array}{ccc}
U_{e1} & U_{e2} & U_{e3} \\
U_{\mu1} & U_{\mu2} & U_{\mu3} \\
U_{\tau1} & U_{\tau2} & U_{\tau3}
\end{array}\right)
\left(\begin{array}{c} \nu_1 \\ \nu_2 \\ \nu_3 \end{array}\right) \;,
\end{equation}
where the elements $U_{\alpha i}$ can be expressed in terms of 3 angles
and one (three) phase(s) for Dirac (Majorana) neutrinos. It is
convenient to introduce the notation
\begin{equation}
P_{\alpha j} \equiv |U_{\alpha j}|^2 \,.
\label{notation}
\end{equation}
The unitarity constraint on the $P_{\alpha j}$ is
\begin{equation}
\sum_j P_{\alpha j} = 1 \,.
\label{unitarity}
\end{equation}
Without loss of generality we take the mass eigenvalues to be real and
positive in the following discussion, and order them as
\begin{equation}
m_3 > m_2 > m_1 > 0 \,.
\label{order}
\end{equation}
The atmospheric neutrino oscillations have a
substantially larger $\delta m^2$ than the solar oscillations, so
we define
\begin{eqnarray}
\delta m_{\rm atm}^2 &=& m_3^2 - m_1^2 \,,
\label{dm_atm} \\
\delta m_{\rm sun}^2 &=& m_2^2 - m_1^2 {\rm~~or~~} m_3^2 - m_2^2 \,,
\label{dm_sun}
\end{eqnarray}
where the two possibilities for $\delta m^2_{sun}$ correspond to the two
possible mass-squared scenarios shown in Fig.~\ref{fig1}. By convention
we take $\delta m^2_{atm}$ and $\delta m^2_{sun}$ to be positive since
negative signs can be absorbed into the mixing matrix $U$. The relation
\begin{equation}
m_3 = \sqrt{m_1^2 + \delta m^2_{atm}}\, \ge \,\sqrt{\delta m^2_{atm}}\,,
\label{m3}
\end{equation}
gives a lower limit on the largest mass via
%
%
%
\begin{equation}
m_3 - m_1 = {\delta m^2_{atm} \over m_3 + m_1}
\le {\delta m^2_{atm} \over m_3}
\le \sqrt{\delta m^2_{atm}} \, .
\label{dm31}
\end{equation}
%
These equations apply equally to the $\delta m^2_{sun}
= m_2^2-m_1^2$ or $\delta m^2_{sun} = m_3^2-m_2^2$ cases.
Thus the three masses are all degenerate to within (see footnote 1)
\begin{equation}
|m_i - m_j| \leq \sqrt{\delta m^2_{atm}} \simeq 0.02 - 0.08 {\rm~eV} \,,
\label{degen}
\end{equation}
and are therefore indistinguishable in the tritium endpoint measurement.

The tritium endpoint constraint is
\begin{equation}
\sum_j P_{ej} m_j^2 < m_\beta^2 \,.
\label{endpoint}
\end{equation}
Using Eq.~(\ref{unitarity}) and $m_j^2 = m_3^2 - \delta m_{3j}^2$,
Eq.~(\ref{endpoint}) can be expressed as
\begin{equation}
m_3^2 - P_{e1} \delta m_{31}^2 - P_{e2} \delta m_{32}^2
< m_\beta^2 \,.
\label{bound}
\end{equation}
It immediately follows from Eqs.~(\ref{order}) and (\ref{bound}) and the
unitarity condition $P_{e1}+P_{e2} \le 1$ that
\begin{equation}
m_3 < \sqrt{m_\beta^2 + \delta m^2_{atm}} \,,
\label{bound2}
\end{equation}
and since $\delta m^2_{atm}$ is small compared to $m_\beta^2$,
Eqs.~(\ref{m3}) and (\ref{bound2}) constrain $m_3$ to the range
\begin{equation}
\sqrt{\delta m^2_{atm}} \leq m_3 < m_\beta \,,
\label{m3max3}
\end{equation}
which numerically bounds $m_3$ to the range
\begin{equation}
0.02\,{\rm~eV} \lsim m_3 \lsim 4.4 {\rm~eV} \,.
\end{equation}
Also, we have
\begin{equation}
\sqrt{\delta m^2_{sun}} \leq m_2 \leq m_3 \,.
\end{equation}
In hierarchical mass spectra in which $m_1 \simeq m_2 \ll m_3$ or
$m_1 \ll m_2 \simeq m_3$, one obtains the equality
\begin{equation}
m_3 \simeq \sqrt{\delta m^2_{atm}} \simeq 0.02 - 0.08 {\rm~eV} \,.
\label{stringent}
\end{equation}
The effective mass-squared of a flavor state is
\begin{equation}
m_{\nu_\alpha}^2 = \sum_j P_{\alpha j} m_j^2 \,.
\end{equation}
When combined with Eqs.~(\ref{unitarity}) and (\ref{order}), we obtain
\begin{equation}
m_{\nu_\alpha} \alt m_3 \,,
\label{nolabel}
\end{equation}
for $\alpha = e,\ \mu$, or $\tau$.

This derivation of mass bounds is independent of the values of the
mixing-matrix; unitarity of the mixing matrix is sufficient.  Values or
bounds for the $\Pej$ inferred from solar and atmospheric neutrino data
may be input, but as seen in (\ref{bound}), they affect the bound only
to order ($\dmatm/\mb$).  The tritium endpoint constraint in
Eq.~(\ref{endpoint}) is also quite general, given the sizes of the
oscillation mass scales $\delta m^2_{atm}$ and $\delta m^2_{sun}$;
within the precision of the tritium experiment, all three mass
eigenstates would appear at the same place in the tritium decay
spectrum. If the common mass is too large to have sufficient phase space
to contribute to tritium decay, then tritium decay could not occur at
all; since tritium decay is observed, the limit applies in all possible
cases.

The above results are valid for Majorana or Dirac neutrinos, because
neutrino oscillations and beta decay are determined by the mass--squared
eigenvalues of the Hermitian matrix $MM^{\dag}$.  However, neutrinoless
double--beta decay measurements offer the possibility of further
limiting elements of the mass matrix itself for the case of Majorana
neutrinos. For Majorana neutrinos, non-observation of neutrinoless
double-beta decay\cite{klap} gives a bound of
\begin{equation}
|M_{ee}| = |\sum_j m_j P_{ej} e^{i\phi_j}| < 0.46{\rm\ eV} \,,
\label{bb}
\end{equation}
where the $\phi_j$ are possible phases from the $U_{ej}$ or the entries
in the diagonal mass matrix. Since $m_3$ is the largest mass eigenvalue,
unitarity leads to the condition
\begin{equation}
|\sum_j m_j P_{ej} e^{i\phi_j}| \le m_3 \,.
\label{m3sum}
\end{equation}
If $m_1 < 0.38$~eV, then from Eq.~(\ref{degen}), $m_3 < 0.46$~eV; it
then follows from Eq.~(\ref{m3sum}) that
Eq.~(\ref{bb}) is satisfied for all values of the mixing matrix elements
$U_{ej}$ and phases $\phi_j$. On the other hand, if $m_1 \ge 0.38$~eV,
Eq.~(\ref{bb}) may give an additional constraint, depending on the values
of the $m_j$, $U_{ej}$, and $\phi_j$.

These considerations can be extended to the four--neutrino case,
assuming the fourth neutrino is sterile and that neutrino oscillations
also explain the recent LSND results
\footnote{The LSND data\cite{LSND} indicate a mass--squared difference in the
range $0.3$~eV$^2 < \delta m^2_{LSND} < 10$~eV$^2$, although
the KARMEN experiment\cite{KARMEN} rules out part of the LSND allowed region.
}(see, e.g., Ref.~\cite{caldwell}).
Equation~(\ref{endpoint}) may then be written as
\begin{equation}
m_4^2 - \sum_{j<4} P_{ej} \delta m_{4j}^2 < m_\beta^2 \,,
\label{bound3}
\end{equation}
where we assume
\begin{equation}
m_4 > m_3 > m_2 > m_1 > 0 \,.
\label{order2}
\end{equation}
In this case $m_4^2-m_1^2 = \delta m^2_{LSND}$
defines the mass--squared difference that describes the LSND results.
Then from Eqs.~(\ref{bound3}) and (\ref{order2}), and the
unitarity condition $P_{e1}+P_{e2}+P_{e3}~\leq~1$, it follows that
\begin{equation}
\sqrt{\delta m^2_{LSND}} \leq m_4
< \sqrt{ m_\beta^2 + \delta m^2_{LSND} } \lsim 5.4 {\rm~eV} \,,
\label{bound4}
\end{equation}
which numerically bounds $m_4$ to the range
\begin{equation}
0.5\,{\rm~eV} \lsim m_4 \lsim 5.4 {\rm~eV} \,.
\end{equation}

This four-neutrino result is likely to be stable even if more neutrinos exist.
The generalization of Eqs. (\ref{endpoint}), (\ref{order2}), and
(\ref{bound4}) to a more-neutrinos universe is
\begin{equation}
m_4 < (\sum^*_j P_{ej})^{-1/2}\,
\left[ m_\beta^2 + \delta m^2_{41} \sum_{j<4} P_{ej}-\sum^*_{j>4}
P_{ej}\,\delta m^2_{j4}\right]^{1/2}\; ,
\label{boundN}
\end{equation}
where the $^*$ on the sum reminds us to include only neutrino masses
small enough to appear in the phase space near the end point spectrum
of tritium decay.
We know that $\sum^*_j P_{ej}$, which is the overlap of $\nu_e$ with the
mass states kinematically accessible to beta decay, is close to unity
from the universality of the coupling strength in beta decay and in other
weak interaction channels.  Setting $\sum^*_j P_{ej}\simeq 1$ in Eq.~(\ref{boundN})
reveals that if $\delta m^2_{41}$ is $\lsim \delta m^2_{LSND}$, then
$m_4$ still satisfies the bound in Eq.~(\ref{bound4}).

In the same way, it can be shown that Eqs.~(\ref{bound2})--(\ref{stringent})
continue to hold if more than three neutrinos exist, if
$\delta m^2_{31}$ is $\lsim \delta m^2_{atm}$. Then, Eq.~(\ref{nolabel})
follows from universality of the weak coupling in processes which produce
the $\nu_e$, $\nu_{\mu}$, and $\nu_{\tau}$, e.g.,  in beta, muon,
and tau decay, respectively.

The bounds in Eqs.~(\ref{m3max3}), (\ref{stringent}), and (\ref{bound4})
have very different implications for cosmology. The fractional
contribution from neutrinos to the closure density of the universe is
$\Omega_\nu = 0.02 (\sum m_\nu/{\rm eV})h^{-2}_{70}$, where $h_{70}$ is
the Hubble constant in units of 70~km/s/Mpc. With the three--neutrino mass
hierarchy $m_1^2 \ll m_3^2$, neutrino mass makes an insignificant
contribution to hot dark matter. On the other hand,
with the three--neutrino degenerate case or with four neutrinos,
the contribution to hot dark matter may be relevant to
large-scale structure formation \cite{primack}.

In summary, if oscillations of three neutrino flavors explain the
atmospheric and solar neutrino data, then all the differences in masses
must satisfy $|m_i - m_j| \lsim 0.08$~eV. The tritium beta decay endpoint
constraint then leads to $m_{\nu_\alpha} \lsim 4.4$~eV for $\alpha = e,
\mu, \tau$, and in the case of a hierarchical spectrum with one or two
neutrino mass eigenstates much lighter than $m_3$, the upper limit becomes
$m_{\nu_\alpha} \lsim 0.08$~eV. If oscillations of one sterile and three
active neutrinos explain the LSND, atmospheric, and solar data, the
tritium beta decay endpoint
constraint then leads to $m_{\nu_\alpha} \lsim 5.4$~eV for $\alpha = e,
\mu, \tau, s$.

\newpage
{\it\underline{Acknowledgements}.}
We thank Sandip Pakvasa for collaboration on previous related work.
This work was supported in part by the U.S. Department of Energy,
Division of High Energy Physics, under Grants No.~DE-FG02-94ER40817,
No.~DE-FG05-85ER40226, and No.~DE-FG02-95ER40896, and in part by the
University of Wisconsin Research Committee with funds granted by the
Wisconsin Alumni Research Foundation and the Vanderbilt University
Research Council.


\end{document}